# Twitter Big Data as a Resource for Exoskeleton Research: A Large-Scale Dataset of about 140,000 Tweets and 100 Research Questions

**Nirmalya Thakur**

Department of Electrical Engineering and Computer Science, University of Cincinnati, Cincinnati, OH 45221-0030, U.S.A.; thakurna@mail.uc.edu

**Abstract:** The exoskeleton technology has been rapidly advancing in the recent past due to its multitude of applications and diverse use-cases in assisted living, military, healthcare, firefighting, and industry 4.0. The exoskeleton market is projected to increase by multiple times of its current value within the next two years. Therefore, it is crucial to study the degree and trends of user interest, views, opinions, perspectives, attitudes, acceptance, feedback, engagement, buying behavior, and satisfaction, towards exoskeletons, for which the availability of Big Data of conversations about exoskeletons is necessary. The Internet of Everything style of today's living, characterized by people spending more time on the internet than ever before, with a specific focus on social media platforms, holds the potential for the development of such a dataset by the mining of relevant social media conversations. Twitter, one such social media platform, is highly popular amongst all age groups, where the topics found in the conversation paradigms include emerging technologies such as exoskeletons. To address this research challenge, this work makes two scientific contributions to this field. First, it presents an open-access dataset of about 140,000 tweets about exoskeletons that were posted in a 5-year period from May 21, 2017, to May 21, 2022. Second, based on a comprehensive review of the recent works in the fields of Big Data, Natural Language Processing, Information Retrieval, Data Mining, Pattern Recognition, and Artificial Intelligence that may be applied to relevant Twitter data for advancing research, innovation, and discovery in the field of exoskeleton research, a total of 100 Research Questions are presented for researchers to study, analyze, evaluate, ideate, and investigate based on this dataset.

## 1. Introduction

A robotic exoskeleton may be broadly defined as a wearable electromechanical device developed with the primary objective of augmenting the physical performance, stamina, and abilities of the person who wears it [1]. Depending on the specific application for which the exoskeleton would be used, they differ in design, functionality, operation, and necessary maintenance [2]. Exoskeletons can be broadly classified as upper limb exoskeletons [3] and lower limb exoskeletons [4]. In the last few years, exoskeletons have been developed for specific body parts or joints. These include exoskeletons for the knee [5], shoulder [6], elbow [7], ankle [8], waist [9], hip [10], neck [11], spine [12], wrist [13], and index finger [14]. The exoskeleton technology has been rapidly advancing [15] in the recent past on account of its multitude of applications and use cases. Some of the applications of exoskeletons [16-18] include – (1) assisted living: helping older adults as well as people with various forms of disabilities to perform their daily routine tasks independently; (2) military: for augmenting productivity and lowering fatigue; (3) healthcare:

to improve the quality of life of people who have lost one or more of their arms or legs or suffer paralysis in the same; (4) firefighting: to assist firefighters in climbing faster as well as for helping them to lift and carry heavy equipment's; (5) industry 4.0: to increase labor productivity, to assist in the transportation of heavy machinery, and to help workers in labor-intensive tasks. Given these diverse applications of exoskeletons and the projected increase in the same, the exoskeleton market is growing rapidly. It was USD 200 million in 2017 [19], and it is estimated to become USD 1.3 billion by the end of 2024 [20].

With the advent and rapid adoption of the Internet of Everything lifestyle [21], in recent times, people's everyday lives involve interacting with computers and a myriad of technology-based gadgets and devices in multiple ways while using various forms of Internet-based services and applications, a lot more than ever before [22, 23]. In this Internet of Everything era, the use of social media platforms has skyrocketed in the recent past [24]. Social media platforms have evolved and transformed into 'virtual' spaces via which people from all demographics communicate with each other to form their social support systems [25] and develop 'online' interpersonal relationships [26]. Such 'virtual' spaces consist of conversations on diverse topics such as emerging technologies, news, current events, politics, family, relationships, and career opportunities [27]. When such conversations are related to a specific technology, the study of the same can indicate the degree and trends of user interest, views, opinions, perspectives, attitudes, and feedback towards that specific technology [28]. Such inferences from conversations about a technology can serve a wide range of use cases such as the indication of the market potential [29], sales prediction [30], consumer engagement [31], technology acceptance [32], buying behavior [33], and customer satisfaction [34], just to name a few.

Twitter, one such social media platform, used by people of almost all age groups [35], has been rapidly gaining popularity in all parts of the world and is currently the second most visited social media platform [36]. At present, there are about 192 million daily active users on Twitter [20], and approximately 500 million tweets are posted on Twitter every day [37]. Therefore, mining and studying this Big Data of conversations from Twitter has been of significant interest to the research community. In the last few years, there have been several works in the fields of Big Data, Data Mining, and Natural Language Processing related to the development of datasets of Twitter conversations related to different topics, technologies, events, diseases, viruses, etc., such as – movies [38], COVID-19 [39], elections [40], toxic behavior amongst adolescents [41], music [42], natural hazards [43], personality traits [44], civil unrest [45], drug safety [46], climate change [47], hate speech [48], migration patterns [49], conspiracy theories [50], and Inflammatory Bowel Disease [51], just to name a few. Recent studies [52-54] have shown that sharing such data helps in the advancement of research, improves the quality of innovation, supports better investigation, and helps to avoid redundant efforts.

In view of the rapid advances in exoskeleton research in the recent past [15], its applicability for a wide range of use cases for diverse users with diverse needs [3-14], projected increase of the booming exoskeleton market to become USD 1.3 billion by the end of 2024 [20], and limitations in prior works in this field [3-14] that did not focus on mining social media conversations about exoskeletons which could have served as an information resource for studying user interest, views, opinions, perspectives, attitudes, acceptance, and feedback towards exoskeletons, it is crucial to develop a dataset of conversations related to exoskeletons. The work presented in this paper aims to address this research challenge by exploring the intersections of Big Data, Data Mining, Natural Language Processing, Information Retrieval, Internet of Everything, and their interrelated disciplines. It makes the following scientific contributions:

1. It presents a large-scale open-access Twitter dataset of 138,585 tweets (including original tweets, retweets, and replies) about exoskeletons posted on Twitter for a period of 5-years from May 21, 2017, to May 21, 2022. The dataset is available at https://dx.doi.org/10.21227/r5mv-ax79.

2. Based on a comprehensive review of 108 emerging works in these fields, this paper discusses multiple interdisciplinary applications of this dataset and presents a list of 100 research questions for researchers to study, analyze, evaluate, ideate, and investigate based on this dataset.

The rest of this paper is organized as follows. The methodology that was used to develop this dataset is presented in Section 2. Section 3 presents the results. Section 4 discusses multiple interdisciplinary applications of this dataset and presents 100 research questions for researchers to investigate. Section 5 concludes the paper by summarizing the scientific contributions of this research and discussing the scope for future work.

**2. Literature Review**

This section describes the methodology that was followed for the development of this dataset, which is publicly available at https://dx.doi.org/10.21227/r5mv-ax79. The dataset contains the Tweet IDs of 138,585 tweets about exoskeletons that were posted over a 5-year period from May 21, 2017, to May 21, 2022. Here, May 21, 2022, is the most recent date of data collection at the time of submission of this paper, and May 21, 2017, is the earliest date for which tweets could be mined when the process of data collection was started. As a central focus of this work involves developing a Twitter dataset, therefore the privacy policy, developer agreement, and guidelines for content redistribution of Twitter [55,56] were thoroughly studied. The privacy policy of Twitter [55] states – *"Twitter is public and Tweets are immediately viewable and searchable by anyone around the world"*. To add, the Twitter developer agreement [56] defines tweets as *"public data"*. The guidelines for Twitter content redistribution [56] state – "*If you provide Twitter Content to third parties, including downloadable datasets or via an API, you may only distribute Tweet IDs, Direct Message IDs, and/or User IDs (except as described below)*. It also states - *"We also grant special permissions to academic researchers sharing Tweet IDs and User IDs for non-commercial research purposes. Academic researchers are permitted to distribute an unlimited number of Tweet IDs and/or User IDs if they are doing so on behalf of an academic institution and for the sole purpose of non-commercial research."* Therefore, it may be concluded that mining relevant tweets from Twitter to develop a dataset (comprising only Tweet IDs) is in compliance with the privacy policy, developer agreement, and content redistribution guidelines of Twitter (at the time of writing of this paper).

The tweets were collected by using the Search Twitter *operator* [57] in RapidMiner [58] and the Advanced Search feature of the Twitter API. RapidMiner is a data science platform that allows the development, implementation, and testing of various algorithms, processes, and applications in the fields of Big Data, Data Mining, Data Science, Artificial Intelligence, Machine Learning, and their related areas. Every application developed in RapidMiner is known as a *process*, and every RapidMiner *process* comprises one or more *operators*, which represent different operational features of the application. The Search Twitter *operator* works by building a connection with the Twitter API while following the rate limits for accessing Twitter data as per Twitter's policy [59]. This ensures that the Search Twitter operator functions by complying with the Twitter API search policies [60]. To use the Search Twitter *operator*, a *process* was developed in the RapidMiner studio that comprised this *operator*. The Search Twitter *operator* requires mandatory input from the user for the query *field*. Here query represents a keyword or the set of keywords or phrases based on which relevant Tweets (Tweets containing that keyword or keywords or phrases) would be filtered and returned in the form of results.

So, to determine the input for this query field, the previous works [1-19] in the field of exoskeleton technology were studied to determine the most common phrases or set of keywords that are used to refer to the underlining exoskeleton systems. A list of phrases and keywords can be found in these works, which include – *"Exoskeleton-Type Systems"*, *"Upper-Limb Exoskeleton"*, *"Indego Explorer Lower-Limb Exoskeleton"*, *"Rehabilitation Exoskeleton,"* *"Powered Knee Exoskeleton"*, *"Exo4Work Shoulder Exoskeleton,"*, *"Wearable Elbow Exoskeleton"*, *"Powered Ankle Exoskeleton"*, *"Wearable Waist Exoskeleton"*, *"Powered Hip*

*Exoskeleton", "Neck Supporting Exoskeleton", "Elastic Spine Exoskeleton", "Wrist Exoskeleton", "Robotic Exoskeleton", "Finger Exoskeleton", and "Lower Limb Exoskeleton"*. As can be seen from this collection of phrases, the keyword "exoskeleton" always exists in a phrase that is being used to refer to a specific kind of exoskeleton system. Therefore, in the *query* field of the Search Twitter *operator*, only the keyword "exoskeleton" was entered. RapidMiner is not case-sensitive, so it ensured that the results would contain tweets where any form of upper case or lower case combinations have been used to spell the word "exoskeleton" in a tweet. The output of this *process* in RapidMiner that comprised the Search Twitter *operator* consisted of multiple attributes, which are mentioned in Table 1. This RapidMiner *process* was run multiple times, and the results were merged together with the results from the Advanced Search Twitter API to develop the dataset and its associated files.

**Table 1.** Description of the attributes from the results of the RapidMiner *process* that used the Search Twitter *operator*.

| Attribute Name | Description |
|---|---|
| Row no. | Row number of the results |
| Id | ID of the tweet |
| Created-At | Date and time when the tweet was posted |
| From-User | Twitter username of the user who posted the tweet |
| From-User-Id | Twitter User ID of the user who posted the tweet |
| To-User | Twitter username of the user whose tweet was replied to (if the tweet was a reply) in the current tweet |
| To-User-Id | Twitter user ID of the user whose tweet was replied to (if the tweet was a reply) in the current tweet |
| Language | Language of the tweet |
| Source | Source of the tweet to determine if the tweet was posted from an Android source, Twitter website, etc. |
| Text | Complete text of the tweet, including embedded URLs |
| Geo-Location-Latitude | Geo-Location (Latitude) of the user posting the tweet |
| Geo-Location-Longitude | Geo-Location (Longitude) of the user posting the tweet |
| Retweet Count | Retweet count of the tweet |

To ensure compliance with the Privacy Policy, Developer Agreement, and Content Redistribution guidelines of Twitter [55,56] for complete data anonymization, and to comply with the FAIR (Findability, Accessibility, Interoperability, and Reusability) principles for scientific data management [61], multiple data filters were introduced in the RapidMiner *process* to filter out all the attributes from the results, other than the "Id" attribute, that represents the unique Tweet ID for each tweet. After running the process, each time, these Tweet IDs were exported as a .csv file and then converted to a .txt file to facilitate easy hydration of the tweets (explained in Section 3) later on. As the dataset contains close to 140,000 Tweet IDs, to ensure that the process of hydration of the Tweet IDs is easy and less time-consuming, the dataset is divided into seven sets or files (7 .txt files) based on the date ranges of the associated tweets. An overview of these sets is shown in Table 2. It is worth mentioning here that the Twitter API's standard search feature (that was used for the dataset development via the Search Twitter *operator* in RapidMiner) does not return a complete index of all Tweets in a date range. So, it is possible that multiple tweets about exoskeletons posted in the date range of May 21, 2017, to May 21, 2022, were not collected by the Twitter API's standard search and are thus not a part of this dataset. To add, Twitter allows users the option to delete a tweet which would mean that there would be no retrievable Tweet text and other related information (upon hydration) for a Tweet ID of a deleted tweet.

**Table 2.** Characteristics of the Dataset Files with the associated Tweet ID count and Date Range

| Filename | Number of Tweet IDs | Date Range of the Tweets |
|---|---|---|
| Exoskeleton_TweetIDs_Set1.txt | 22945 | July 20, 2021 – May 21, 2022 |
| Exoskeleton_TweetIDs_Set2.txt | 19416 | Dec 1, 2020 – July 19, 2021 |
| Exoskeleton_TweetIDs_Set3.txt | 16673 | April 29, 2020 - Nov 30, 2020 |
| Exoskeleton_TweetIDs_Set4.txt | 16208 | Oct 5, 2019 - Apr 28, 2020 |
| Exoskeleton_TweetIDs_Set5.txt | 17983 | Feb 13, 2019 - Oct 4, 2019 |
| Exoskeleton_TweetIDs_Set6.txt | 34009 | Nov 9, 2017 - Feb 12, 2019 |
| Exoskeleton_TweetIDs_Set7.txt | 11351 | May 21, 2017 - Nov 8, 2017 |

## 3. Results and Discussions

As mentioned in Section 2, this dataset contains Tweet IDs corresponding to 138,585 tweets about exoskeletons posted on Twitter from May 21, 2017, to May 21, 2022. The complete information associated with a tweet, such as the text of a tweet, user name, user ID, timestamp, retweet count, etc., can be obtained from a Tweet ID by following a process known as hydration of Tweet ID [62]. A simple example of this process can be observed by visiting this website [63] and entering any Tweet ID from this dataset. The website would immediately generate the URL of the associated Tweet that can be studied. However, in a realistic scenario repeating this step 138,585 times would be a very difficult process. In view of the same and for processing Twitter datasets, researchers in this field have developed various tools that can hydrate Tweet IDs. Some of the most popular tools include – Hydrator [64], Social Media Mining Toolkit [65], and Twarc [66]. This section comprises two parts. In Section 3.1, the process of hydration in the context of using one of these tools – the Hydrator app, is explained, and a brief discussion about the results that would be obtained thereafter is also presented. Section 3.2 presents a statistical analysis of multimodal components of the Tweets of this dataset.

### *3.1. Hydrating the Dataset – Steps and Associated Results*

In this context, results mean the complete information (text of the tweet, user ID, user name, retweet count, language, tweet URL, source, and other public information related to the tweet) associated with each of the Tweet IDs in this dataset.

The following is the step-by-step process for using the Hydrator app to hydrate this dataset. For the purpose of this discussion, only one file from this dataset – "Exoskeleton_TweetIDs_Set2.txt," is being used.

1. Download and install the desktop version of the Hydrator app from this website [67]. The version that was used for this work is v0.30.0.
2. Click on the "Link Twitter Account" button on the Hydrator app to connect the app to an active Twitter account.
3. Click on the "Add" button to add a new dataset comprising only Tweet IDs (Figure 1). Browse and select the file – "Exoskeleton_TweetIDs_Set2.txt" available on local storage.
4. If the file upload is successful, the Hydrator app will show the total number of Tweet IDs present in the file. For this file - "Exoskeleton_TweetIDs_Set2.txt", the app would show the Number of Tweet IDs as 19,415.
5. Provide details for the respective fields: Title, Creator, Publisher, and URL in the app, and click on "Add Dataset" to add this dataset to the app.
6. The app would automatically redirect to the "Datasets" tab. Click on the "Start" button to start hydrating the Tweet IDs. During the hydration process, the progress indicator would increase, indicating the number of Tweet IDs that have been successfully hydrated and the number of Tweet IDs that are pending hydration.
7. After the hydration process ends, a .jsonl file would be generated by the app that the user can choose to save. The app would also display a "CSV" button in place of the "Start" button. Clicking on this "CSV" button would generate a .csv file with detailed information about the tweets, which would include the text of the tweet, user ID, user name, retweet count, language, tweet URL, source, and other public information related to the tweet.

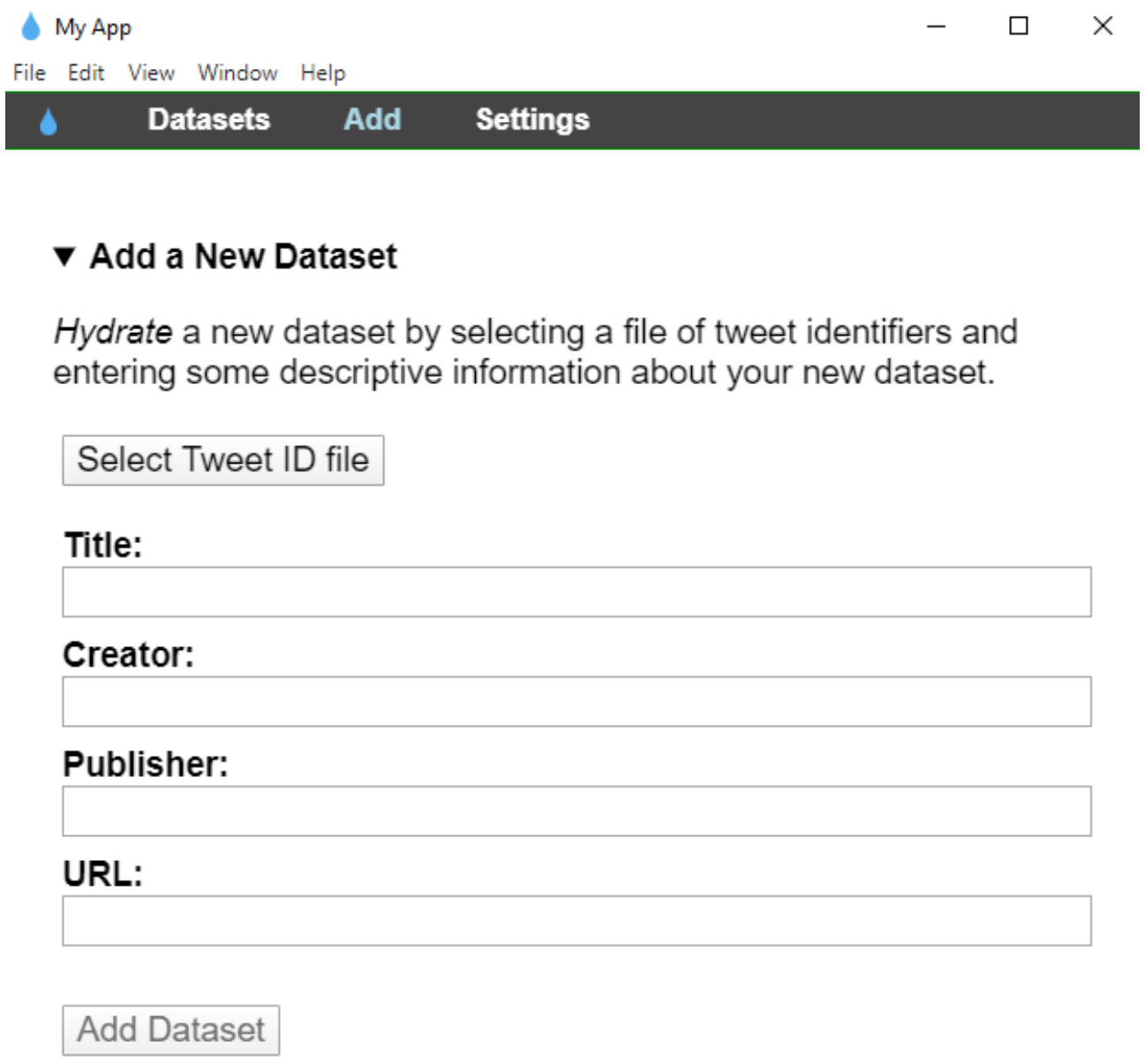

**Figure 1.** Screenshot from the Hydrator app for the dataset upload step

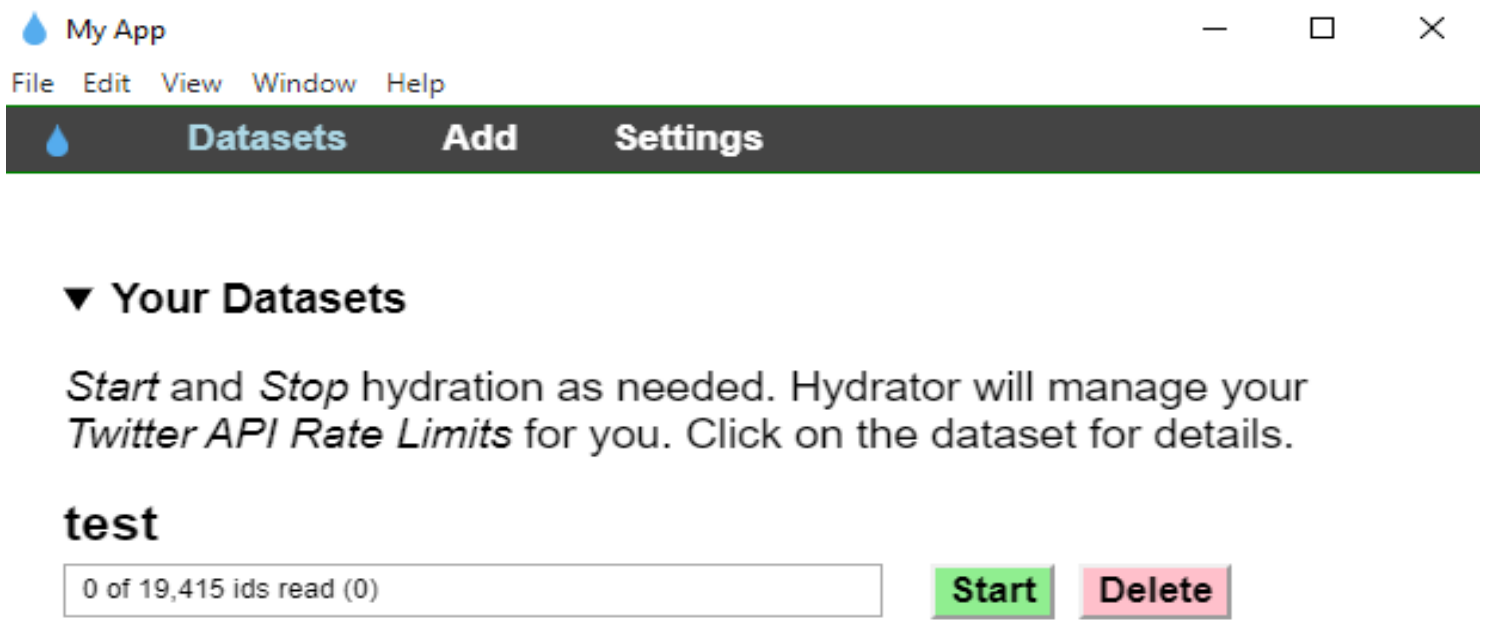

**Figure 2.** Screenshot from the Hydrator app for the dataset processing step

Even though the above steps are discussed in the context of using the "Exoskeleton_TweetIDs_Set2.txt" file from the dataset, the same steps can be repeated for hydrating all the other dataset files. However, it is to be noted that Hydrator functions by connecting with the Twitter API and is in compliance with the Twitter API rate limits [59]. So, when the app reaches the 15-minute or 3-hour or any similar usage limit, hydrating any additional dataset files might not be possible in that 15-minute or 3-hour or a similar usage criteria/window.

### *3.2. Statistical Analysis of the Dataset*

This section presents the findings of a comprehensive statistical analysis of this dataset which was performed using RapidMiner [58]. RapidMiner has an in-built statistical analysis function that can be used for studying multimodal characteristics of datasets. The characteristics of these Tweets that were studied include - language, embedded URLs, retweet count, favorite count, length of the tweets, dates when these tweets were posted, and the number of users who posted these tweets. This analysis was performed on July 9, 2022 (the most recent date at the time of re-submission of this paper after the first round of review). Prior to performing this analysis, the Tweet IDs needed to be hydrated, and the Hydrator app [64] was used to Hydrate this dataset (Section 3.1). The Hydrator app reported that 7% of the tweets were deleted at the time of Hydration. Therefore, this analysis is presented from the remaining set of tweets. However, it is worth mentioning here that these specific characteristics could be studied only for those tweets which presented the relevant data. For instance, several tweets were observed to be just images and videos of different types of exoskeletons and retweets of the same. For such tweets, characteristics such as language analysis, embedded URL identification, and computing the length of the tweet did not return any results. A total of 49 different languages were detected in the set of tweets in this dataset. The languages were represented as per the ISO 639 codes [68] in the results that were obtained in RapidMiner. ISO 639 [68] is a standardized nomenclature used to classify languages. Each language is assigned a two-letter (639-1) and three-letter (639-2 and 639-3) lowercase abbreviation, amended in later versions of the nomenclature. These language codes were converted to the actual languages for a seamless understanding of the results. The results are presented in Table 3, along with the exact count and the percentage of tweets posted in each language. As can be seen from Table 3, English was the most common language that was used for tweeting about exoskeletons from May 21, 2017 to May 21, 2022.

**Table 3.** Analysis of the different languages represented in the Tweets present in this dataset

| Language | Language Code | Absolute Count | Percentage |
|---|---|---|---|
| English | en | 71585 | 0.904296308 |
| Danish | da | 1085 | 0.013706244 |
| Tagalog | tl | 926 | 0.011697679 |
| Spanish | es | 774 | 0.009777542 |
| Indonesian | in | 419 | 0.00529301 |
| Japanese | ja | 288 | 0.003638155 |
| French | fr | 267 | 0.003372873 |
| German | de | 206 | 0.002602292 |
| Hungarian | hu | 197 | 0.002488599 |
| Czech | cs | 192 | 0.002425437 |
| Catalan | ca | 183 | 0.002311744 |
| Romanian | ro | 181 | 0.002286479 |
| Turkish | tr | 119 | 0.001503265 |
| Estonian | et | 117 | 0.001478001 |
| Portuguese | pt | 110 | 0.001389573 |
| Finnish | fi | 105 | 0.001326411 |
| Basque | eu | 96 | 0.001212718 |
| Dutch | nl | 89 | 0.001124291 |
| Slovenian | sl | 84 | 0.001061129 |
| Russian | ru | 65 | 8.21E-04 |
| Thai | th | 50 | 6.32E-04 |
| Haitian | ht | 43 | 5.43E-04 |
| Italian | it | 36 | 4.55E-04 |
| Arabic | ar | 31 | 3.92E-04 |
| Lithuanian | lt | 19 | 2.40E-04 |
| Swedish | sv | 18 | 2.27E-04 |
| Polish | pl | 16 | 2.02E-04 |
| Papiamentu | qst | 13 | 1.64E-04 |
| Korean | ko | 11 | 1.39E-04 |
| Kunstsprachen | art | 8 | 1.01E-04 |
| Chinese | zh | 8 | 1.01E-04 |
| Greek | el | 6 | 7.58E-05 |
| Vietnamese | vi | 6 | 7.58E-05 |
| Hindi | hi | 5 | 6.32E-05 |
| Icelandic | is | 5 | 6.32E-05 |
| Norwegian | no | 4 | 5.05E-05 |
| Persian | fa | 3 | 3.79E-05 |
| Hebrew | iw | 3 | 3.79E-05 |
| Welsh | cy | 2 | 2.53E-05 |
| Latvian | lv | 2 | 2.53E-05 |
| Malayalam | ml | 2 | 2.53E-05 |
| Urdu | ur | 2 | 2.53E-05 |
| Bulgarian | bg | 1 | 1.26E-05 |
| Gujrati | gu | 1 | 1.26E-05 |
| Armenian | hy | 1 | 1.26E-05 |
| Georgian | ka | 1 | 1.26E-05 |
| Burmese | my | 1 | 1.26E-05 |
| Tamil | ta | 1 | 1.26E-05 |
| Ukrainian | uk | 1 | 1.26E-05 |

The analysis of the URLs embedded in the tweets is presented next. All the tweets did not have an URL embedded in the tweet text. Out of the tweets that had an URL embedded in the text, a total of 23,831 different URLs were detected. To avoid presenting a table with 23,831 rows, the top 50 URLs in terms of the number of times of occurrence are represented in Table 4. As can be seen from Table 4, the top 2 embedded URLs pointed to the news published by BBC News, a business division of the British Broadcasting Corporation (BBC), about a mind-controlled exoskeleton.

**Table 4.** Analysis of the different URLs embedded in the Tweets present in this dataset

| URL | Absolute Count | Percentage |
|---|---|---|
| https://www.bbc.co.uk/news/health-49907356 | 268 | 0.008193464 |
| https://www.bbc.com/news/health-49907356 | 201 | 0.006145098 |
| http://smarturl.it/xu3k1p | 58 | 0.001773212 |
| https://ehikioya.com/product/wearable-chairless-chair/ | 54 | 0.001650922 |
| http://bit.ly/2tkwtak | 48 | 0.001467486 |
| https://ift.tt/2raBxMq | 47 | 0.001436913 |
| https://ift.tt/2J7R2N4 | 46 | 0.001406341 |
| https://www.theguardian.com/world/2019/oct/04/paralysed-man-walks-using-mind-controlled-exoskeleton?CMP=share_btn_tw | 45 | 0.001375768 |
| http://youtu.be/0z0i8GzfyXs?a | 40 | 0.001222905 |
| http://youtu.be/qTxxwLWsMoA?a | 40 | 0.001222905 |
| https://www.caradvice.com.au/674659/ford-expands-trial-of-exoskeleton-vest/ | 40 | 0.001222905 |
| http://bit.ly/1fZbb79 | 39 | 0.001192332 |
| https://www.bbc.co.uk/news/technology-44628872 | 38 | 0.00116176 |
| http://www.bbc.co.uk/news/technology-44628872 | 36 | 0.001100615 |
| http://youtu.be/OiAVTz5BbZQ?a | 36 | 0.001100615 |
| https://techcrunch.com/2018/03/29/roam-debuts-a-robotic-exoskeleton-for-skiers/ | 36 | 0.001100615 |
| https://prescient.info/xLHKDINg/ | 34 | 0.001039469 |
| https://spectrum.ieee.org/the-human-os/biomedical/devices/cyberdynes-medical-exoskeleton-strides-to-fda-approval | 34 | 0.001039469 |
| http://bit.ly/2eNowmn | 33 | 0.001008897 |
| https://cbsn.ws/31Ou-fyB?fbclid=IwAR1t_TSzmgKxZ73ASQ5Uc9nWoQSJoQyS45bwieBCAwKL00nCHXAJeebJMrw | 33 | 0.001008897 |
| https://mashable.com/2018/02/06/ford-exoskeleton-factory-ekso/ | 33 | 0.001008897 |
| https://www.bbc.co.uk/news/av/technology-47319932/exoskeleton-helps-people-with-paralysis-to-walk | 33 | 0.001008897 |
| http://toastradio.com/np/71C894A884189507 | 32 | 0.000978324 |
| https://prescient.info/xdtmhojB/ | 31 | 0.000947751 |
| https://twitter.com/BadMedicalTakes/status/1499867782934142977 | 31 | 0.000947751 |
| http://ift.tt/2CluPrh | 30 | 0.000917179 |
| https://prescient.info/xhzMlhIZ/ | 30 | 0.000917179 |
| https://www.engadget.com/2017/07/06/russian-exoskeleton-suit-turns-soldiers-into-stormtroopers/ | 30 | 0.000917179 |
| https://youtu.be/zLWuHo63C8k | 29 | 0.000886606 |
| https://futurism.com/amazons-alexa-helps-this-exoskeleton-respond-to-spoken-instructions/ | 28 | 0.000856034 |
| http://wknc.org/listen | 27 | 0.000825461 |
| https://tunein.com/radio/NOFM-s184668/ | 27 | 0.000825461 |
| http://twinybots.ch https://www.forbes.com/sites/servicenow/2020/06/11/ai-is-the-brains-exoskeleton/ | 26 | 0.000794888 |
| https://ift.tt/2F4GCzy | 26 | 0.000794888 |
| https://www.bbc.com/news/technology-44628872 | 26 | 0.000794888 |
| https://www.forbes.com/sites/servicenow/2020/06/11/ai-is-the-brains-exoskeleton/ | 26 | 0.000794888 |
| https://www.reuters.com/lifestyle/father-builds-exoskeleton-help-wheelchair-bound-son-walk-2021-07-26/ | 26 | 0.000794888 |
| http://ift.tt/2usZoYj | 25 | 0.000764316 |
| https://design-milk.com/morpheus-hotel-zaha-hadid-architects-worlds-first-high-rise-exoskeleton/ | 25 | 0.000764316 |
| https://wp.me/p8RPw8-eF | 25 | 0.000764316 |
| https://www.theguardian.com/world/2019/oct/04/paralysed-man-walks-using-mind-controlled-exoskeleton | 25 | 0.000764316 |
| http://wp.me/p5QmfL-158 | 24 | 0.000733743 |
| https://www.cnn.com/2019/10/04/health/paralyzed-man-robotic-suit-intl-scli/index.html | 23 | 0.00070317 |
| https://youtu.be/qTxxwLWsMoA | 23 | 0.00070317 |
| http://ift.tt/2uBXYJI | 22 | 0.000672598 |
| https://apple.news/ApW9EjskUS5Wij9-qUhlzEQ | 22 | 0.000672598 |
| https://futurism.com/ford-pilots-new-exoskeleton-lessen-worker-fatigue/ | 22 | 0.000672598 |
| https://ift.tt/2sRBFkC | 22 | 0.000672598 |
| https://twitter.com/blaireerskine/status/1329844449426468865 | 22 | 0.000672598 |
| https://scout.com/military/warrior/Article/Army-Tests-New-Super-Soldier-Exoskeleton-111085386 | 21 | 0.000642025 |

The analysis of the retweet count of all the tweets showed that the highest and lowest number of retweets received by a tweet were 2624 and 0. The average retweet count of each tweet was computed to be 1.127. Upon studying the favorite count of all the tweets, it was observed that, on average, each tweet received a favorite count of 6.915. The highest favorite count of a tweet was observed to be 90740, while the lowest favorite count was observed to be 0 for multiple tweets. Thereafter, the length of the tweets was studied in terms of character counts. For those tweets that had an embedded URL and users tagged, the length of the URL and length of the Twitter user ID(s) of the tagged users were added to the character count of any text present in those respective tweets. Based on this approach, the maximum character count of a tweet was observed to be 949. This specific tweet had 43 users tagged in the tweet, which attributed to its unusually high character count. The lowest character count for a tweet was observed to be 2, and the average character count of all the tweets was observed to be 128.054. Even though the dataset presents tweets posted over a period of 5 years, it was observed that on certain dates, a high number of tweets about exoskeletons were recorded. Table 5 presents the list of 21 specific dates from this dataset when 100 or more than 100 tweets were posted about exoskeletons over a 24-hour period. As can be seen from Table 5, on October 4, 2019, the maximum number of Tweets about exoskeleton were posted in the last 5 years. It was also on the same date that BBC news published the news articles (top 2 URLs in Table 4) about the mind-controlled exoskeleton. This seems to indicate that was a direct correlation between the BBC news posting about this new exoskeleton and the number of tweets about exoskeletons rapidly increasing over a 24-hour period.

**Table 5.** List of specific dates from this dataset when more than 100 tweets about exoskeletons were posted in a single day

| Date | Number of Tweets |
|---|---|
| Oct 04 2019 | 1014 |
| Mar 30 2018 | 214 |
| Nov 10 2017 | 203 |
| Feb 24 2019 | 194 |
| Jul 08 2018 | 188 |
| Oct 07 2019 | 184 |
| Jul 06 2017 | 162 |
| Oct 06 2019 | 160 |
| Nov 22 2019 | 151 |
| Mar 29 2018 | 137 |
| Nov 24 2017 | 113 |
| Nov 29 2018 | 111 |
| Jun 23 2017 | 109 |
| Sep 27 2017 | 109 |
| Dec 14 2020 | 108 |
| Nov 11 2017 | 107 |
| Feb 25 2019 | 104 |
| Jan 09 2019 | 104 |
| Nov 23 2019 | 103 |
| Nov 24 2019 | 101 |
| Nov 28 2017 | 100 |

The analysis of User IDs showed that a total of 73,345 distinct User IDs are present. Or in other words, a total of 73,345 Twitter users accounted for all the tweets about exoskeletons present in this dataset. The highest and lowest number of tweets about exoskeletons posted by a single user was observed to be 3927 and 1, respectively.

## 4. Potential Applications and Research Directions

This comprehensive dataset that consists of Tweet IDs of 138,585 tweets about exoskeletons that were posted over a 5-year period from May 21, 2017, to May 21, 2022, is expected to have multiple interdisciplinary applications related to the advancement of research, innovation, and discovery in the field of exoskeleton technology. Some of these applications could include interpretation and analysis of the degree and trends of user interests, perspectives, opinions, reviews, attitudes, feedback, engagement, acceptance, buying behavior, and satisfaction, related to different exoskeletons used by diverse users for different use cases. To further support the same, based on a comprehensive review of emerging works in the fields of Big Data, Data Mining, Natural Language Processing, Information Retrieval, Pattern Recognition, and Artificial Intelligence that may be applied to relevant Twitter data for advancing research, innovation, and discovery the field of exoskeleton research, a total of 100 Research Questions (RQ's) are presented for researchers to study, analyze, evaluate, ideate, and investigate based on this dataset. This section is divided into two parts. In Section 4.1, this list of 100 RQ's is presented. As can be seen from Section 4.1, each of these RQ's focuses on different problem statements and/or research directions. So, the underlining procedure for the development and implementation process is going to be different for investigating each of these RQ's. Furthermore, all the recent works (Section 4.1) that have been reviewed to develop these RQ's involved the usage of different programming languages such Python, R, C++, JavaScript, Java, Ruby etc., and research platforms/tools such as RapidMiner, Tableau, Visual Basic, .NET etc., for the development of the associated software framework. So, preparing a common code that can be directly applied to investigate all these RQ's is not possible. However, to facilitate easier investigation and a starting point for studying some of these RQ's, a step-by-step methodology in RapidMiner is described in Section 4.2.

### *4.1. List of 100 Research Questions related to Exoskeletons*

These RQ's are presented as follows:

RQ1. Sentiment analysis [69] of these tweets would help to identify the positive, negative, and neutral sentiments associated with these conversations about exoskeletons on Twitter.

RQ2. Deep learning may be used to identify the emotional state of the users in terms of the basic emotional responses - fear, anger, joy, sadness, disgust, and surprise [70], at the time of posting of these respective tweets.

RQ3. Aspect-based sentiment analysis, along with tokenization and lemmatization of these tweets [71], may be performed to identify the specific aspects or subject matters related to exoskeletons to which certain specific sentiments or emotional responses are associated.

RQ4. Studying the trends in tweet counts [72] and the associated sentiments and emotional responses to detect any correlations between the two.

RQ5. Studying the word count of each of these tweets [73] to determine if any correlation exists between the word count and the associated sentiments and emotional responses towards different exoskeleton products.

RQ6. Investigating whether the number of replies or retweets of a tweet [74] posted by a company to share the news about a new kind of exoskeleton could have a correlation with the influence and follower metrics of the company's Twitter profile.

RQ7. Detecting popular tweets [75] related to exoskeletons and studying the subject matters and aspects mentioned in those tweets by performing tokenization and lemmatization.

RQ8. Detecting sarcasm [76] related to exoskeletons and studying for any correlation of sarcasm with the sentiment or emotional response associated with the respective tweets.

RQ9. Identifying the commonly used hashtags associated with tweets related to exoskeletons and detecting the sentiment related to these respective hashtags [77].

RQ10. Analyzing the tweets made by users of exoskeletons to detect the diverse user personas [78] and their associated perspectives, experiences, opinions, and feedback about exoskeletons.

RQ11. Studying trending discussions [79] on Twitter related to exoskeletons and using machine learning to detect these trends in real-time.

RQ12. Studying the trends in sentiments and emotional responses [80] associated with exoskeletons to track if there is any correlation of the same with the trends in exoskeleton sales or its market potential.

RQ13. Investigating for any interdependence between tweets about specific exoskeleton products and sales [81] of those specific exoskeleton products.

RQ14. Performing topic modeling [82] of these respective tweets to interpret the associated communication as news, recommendation, discussion, feedback, perspective, opinion, etc., related to different kinds of exoskeletons.

RQ15. Investigating retweeting patterns of tweets [83] to determine the interest in certain topics related to exoskeletons expressed in the respective tweets.

RQ16. Studying the tweeting patterns and content of the tweets by various companies or manufacturers of exoskeletons to understand their audience management methodologies which include targeting different audiences, concealing subjects, and maintaining authenticity [84].

RQ17. Detecting the Point of Interest (P.O.I.) of a tweet [85], which presents high-level location information about a place, to understand the location-specific opinions, perspectives, or attitudes of the public towards exoskeleton technology.

RQ18. Developing a personalized tweet recommendation system [86] that would present the latest developments in exoskeleton technology, including exoskeletons available for purchase to potential user groups, which may include the elderly, disabled, handicapped, etc.

RQ19. Performing a case study on different kinds of machine learning classifiers [87] to develop sentiment analysis approaches [88] to deduce the best machine learning classifier in terms of performance characteristics for sentiment analysis of tweets related to exoskeleton technology.

RQ20. Performing semantic analysis of the content of each tweet as per the methodology discussed in [89] to determine if any political leaders have influenced the sale or public opinion or perspectives towards a specific kind of exoskeleton.

RQ21. Analysis of tweets to determine the emergence of exoskeletons [90] in different fields such as healthcare and medicine.

RQ22. Studying the mentions of exoskeleton companies in tweets to determine the patterns of customer engagement [91] with each of these companies.

RQ23. Investigating tweets to determine the global [92] and region-specific [93] reason/drive centered around the purchase or use of specific exoskeletons.

RQ24. Development of a tweet ranking model to present important tweets [94] to potential end-users or customers of exoskeletons based on their specific needs or interests.

RQ25. Determining how official accounts on Twitter play a role in the propagation and correction of online rumors related to exoskeletons in different geographic locations [95].

RQ26. Studying the semantics of the tweets to determine how user diversities such as gender differences [96,97] may impact the tweeting patterns and content of tweets centered around exoskeletons, including their usage and needs.

RQ27. Performing content value analysis [98] of tweets to filter the most relevant and least relevant tweets [99] involving exoskeletons.

RQ28. Developing an approach to determine the audience size [100] of any potential tweet related to a specific exoskeleton that might be helpful for the consumer outreach of exoskeleton companies and manufacturers.

RQ29. Application of the gratification theory [101] on these tweets to deduce the factors that gratify users related to different use-cases of exoskeletons.

RQ30. Performing a study on the tweets to investigate the role of news organizations, including regional media, local media, national media, and broadcast news agencies, in the dissemination of the latest developments [102] in the field of exoskeleton technology.

RQ31. Determining the occupation of potential end-users of exoskeletons from their tweets [103], which may be helpful for companies and/or manufacturers to develop or improvise exoskeletons to better assist these end-users in their respective professions.

RQ32. Implementation of the TCV-Rank summarization technique for generating online summaries and historical summaries related to tweets [104] about exoskeletons posted from different geographic regions.
RQ33. Implementation of the TURank (Twitter User Rank) algorithm [105] to find authoritative Twitter users who post tweets related to exoskeletons.
RQ34. Studying the trends of entity linking [106] in tweets about upcoming exoskeletons for different use-cases.
RQ35. Investigating the impact of following clusters of exoskeleton users or exoskeleton companies [107] on the ideologies of the Twitter user(s) over exoskeletons.
RQ36. Analysis of the tweets centered around specific hashtags [108] related to exoskeletons to analyze the tweeting trends and replies related to these hashtags.
RQ37. Implementation of the HybridSeg approach [109] to find the optimal segmentation of tweets related to exoskeletons for improving segmentation quality as well as for exploring applications of this approach for named entity recognition.
RQ38. Interpretation of the use of Twitter by companies or organizations in the exoskeleton industry to examine brand attributes (both product-related and non-product-related) and their relation to Twitter's key engagement features (Reply, Retweet, Favorite) [110].
RQ39. Development of an approach by application of the Latent Dirichlet Allocation (LDA) model as proposed in [111] to deduce the information credibility related to exoskeleton-based tweets originating from different sources.
RQ40. Studying tweets to detect and predict any potential conspiracy theories [112,113] related to emerging developments in the field of exoskeletons or any specific exoskeleton-based product.
RQ41. Implementation of the Tweet2Vec method [114] for learning tweet embeddings using character-level CNN-LSTM encoder-decoder for efficient categorization of tweets centered around exoskeleton technologies in general or related to any specific exoskeleton technology.
RQ42. Implementation of the Self-Exciting Point Process Model for Predicting Tweet Popularity (SEISMIC) model [115] to predict the popularity of tweets related to exoskeletons.
RQ43. Studying the geographic diffusion patterns in terms of random, local, and information brokerage of the information contained in a specific tweet [116] related to exoskeletons and their diverse use cases.
RQ44. Performing tweet wikification [117] to identify different concepts mentioned in a tweet to link these concepts to existing concepts about exoskeletons present in a knowledge base, such as Wikipedia.
RQ45. Detecting spam accounts [118] and social spam [119] on Twitter that may be the source of spam related to exoskeleton-based information expressed in tweets.
RQ46. Development of an approach similar to the work in [120] for detection of complaints related to specific exoskeleton technologies.
RQ47. Predicting the cost [121, 122] of planned and expected developments in existing exoskeletons based on drawing insights from the tweets about these developments.
RQ48. Application of the approach proposed in [123] to detect the patterns of emojis present in information-based tweets about exoskeletons for the analysis of the relationships between plain texts and emojis usage in such tweets.
RQ49. Tracking and investigating the usage of multiple emojis expressed in the tweets related to different use cases of exoskeletons for investigating the associated sentiment [124,125], performing tweet classification [126], user verification [127], irony detection [128], and trust modeling [129].
RQ50. Detection of fake users [130] who post fake news [131] about exoskeletons on Twitter.

RQ51. Investigating the effect of tweeting about research papers [132] on exoskeletons on the downloads and citations of these respective papers.
RQ52. Determining the social identities [133] of diverse users of exoskeletons based on the content and context of their tweets.
RQ53. Studying the relevance of a tweet [134] about a specific exoskeleton based on the hyperlinked documents in the same.
RQ54. Investigating how exoskeleton companies and/or manufacturers use tagging [135] on Twitter for audience engagement and retention.
RQ55. Performing stance detection [136] towards exoskeletons by analyzing the tweets posted by its users.
RQ56. Interpretation of satire [137] in the context of tweets about new and upcoming exoskeleton technologies.
RQ57. Predicting the age of existing users or potential users of exoskeletons from their tweets [138] to personalize the exoskeletons as per the age-specific needs.
RQ58. Investigating the selective attention over different entities expressed in any tweet pertaining to exoskeletons, as per the methodology proposed in [139].
RQ59. Studying the paradigms of readability in tweets posted by users of exoskeletons to interpret the degrees of engagement [140].
RQ60. Deducing the best time to tweet [141] any information related to exoskeletons that might be helpful for the sales and marketing team of exoskeleton companies and/or manufacturers.
RQ61. Tracking repliers and retweeters of tweets [142] about improvisations in existing exoskeletons posted by exoskeleton companies to detect degrees of intimacy with the target audience.
RQ62. Detecting the number of tweets [143] related to exoskeletons from a geographic area that could be helpful in understanding the associated needs or public perceptions of a specific exoskeleton-based technology available or marketed in that area.
RQ63. Analyzing the multimodal factors that are associated with the retweet of any tweet [144] communicating news about exoskeletons.
RQ64. Using the concept of knowledge graphs for tweet summarization for effectiveness in obtaining useful information [145] related to exoskeleton technologies on Twitter.
RQ65. Recommendation of specific hashtags [146] related to exoskeletons to Twitter users who could be potential users of exoskeletons.
RQ66. Performing contextualization of tweets [147] related to exoskeletons based on hashtags performance prediction and multi-document summarization.
RQ67. Assigning value to tweets related to specific use cases of exoskeletons based on the approach proposed in [148] to compute the worth of the underlining tweets.
RQ68. Deducing the number of followers of exoskeleton companies from their tweets [149] to determine their customer base.
RQ69. Studying tweets for detection of suggestions and classifications of suggestions [150] related to existing and/or emerging technologies associated with exoskeletons.
RQ70. Studying tweets to interpret any forms of discrimination [151] faced by existing or potential users of exoskeletons.
RQ71. Implementation of the iFACT framework [152] on tweets associated with exoskeletons to identify, assess, and evaluate the underlying factual information mentioned in the tweets.
RQ72. Implementation of the SEDTWik framework [153] for segment-based detection of any kinds of events from tweets that focus on the use of exoskeletons by diverse user groups.

RQ73. Developing an approach as per [154] for followee recommendation to existing and/or potential users of exoskeletons based on topic extraction and sentiment analysis from exoskeleton-based tweets.
RQ74. Studying tweets for detecting stress levels and reasons for stress [155] in current or potential users of exoskeletons.
RQ75. Detecting if any tweet about exoskeletons posted by exoskeleton companies can be classified as a "regrettable" tweet [156] so that these companies may delete the tweet to reduce the chances of any potential damage to their reputation.
RQ76. Interpreting diverse activities related to exoskeleton use cases by studying the associated tweets [157] and mapping these activities on pleasure and arousal dimensions using cognitive computing principles.
RQ77. Studying tweets posted by users of exoskeletons to monitor their mental health [158].
RQ78. Detecting deception (both positive and negative deception) from tweets [159] about the use of exoskeletons by specific user groups.
RQ79. Tracking happiness associated with exoskeleton usage in different cities [160] based on studying tweets related to exoskeletons originating from these cities.
RQ80. Identification of hate speech and abusive language in tweets [161] made by unsatisfied customers of exoskeletons.
RQ81. Developing an approach as per [162] to filter out relevant tweets comprising of latest breaking news in the context of exoskeletons.
RQ82. Extracting information from tweets related to exoskeletons to interpret the multimodal forms of purchase intentions [163] in potential users.
RQ83. Inferring shared interests [164] related to exoskeletons based on studying the tweets of both its current and potential users.
RQ84. Modeling public mood in different geographic regions [165] towards new advances in exoskeletons based on semantic analysis of the tweets originating from these respective regions.
RQ85. Implementation of the Categorical Topic Model [166] for extracting categorical topics and emerging issues about exoskeletons from tweets.
RQ86. Using classification approaches to deduce inundation levels [167] in the context of use case scenarios of different exoskeletons by different user groups.
RQ87. Studying the tweets to interpret bias and degrees of the same [168] towards using exoskeletons by potential user groups.
RQ88. Detection, classification, and ranking of trending topics [169] related to conversations about exoskeletons on Twitter.
RQ89. Analyzing tweets related to exoskeletons sold by any specific company to study and predict the changes in that specific company's stock price based on the tweeting patterns [170,171].
RQ90. Performing user characterization [172] from the tweeting patterns of any potential user to develop user personas for personalization of exoskeletons.
RQ91. Studying tweets posted by users of exoskeleton technologies to detect and analyze their feedback and suggestions [173] for possible improvements in the exoskeletons used by these respective users.
RQ92. Application of the Similarity Learning Algorithm (SiLA) as proposed in [174] to identify popular tweets related to current and emerging exoskeletons and their use cases.
RQ93. Implementation of the approach proposed in [175] to study patterns of tweets related to exoskeletons to detect tweets that represent "extreme behavior" on social media.
RQ94. Classifying tweets about specific use cases of exoskeletons as "alarming" and "reassuring" [176] to investigate the views of different user groups.

RQ95. Performing semantic analysis of tweets posted by new users of exoskeletons to detect instances of euphoria or delusion [177] in the context of the use cases mentioned in the underlining tweets.
RQ96. Detecting obesity from tweets [178] made by users of exoskeletons and investigating any potential correlations between obesity and exoskeleton usage.
RQ97. Classifying potential user groups of exoskeletons into communities [179] based on studying their needs expressed in their tweets for the development of specific exoskeletons to meet these community-based needs.
RQ98. Estimating demographic information of exoskeleton users from their tweets [180] to interpret any variation of use cases based on user diversity.
RQ99. Studying the tweets to deduce the perceptions [181] of exoskeletons users about different exoskeleton companies to interpret their buying behavior.
RQ100. Analyzing the tweets to track misinformation and trends in the same [182] about upcoming or existing exoskeletons.

As can be seen from these RQ's, the prior works that were reviewed [69-182] required Tweets (and in some cases, certain characteristics of tweets such as retweet counts, hashtags used in the tweets, date, and time of the tweets, etc. that can be obtained upon Hydration) as a data source. In some of these works, the relevant tweets were mined, and in some of the other works, instead of data mining, a dataset comprising relevant tweets was used. In this dataset, the presence of about 140,000 Tweet IDs corresponding to the same number of tweets upholds the fact that it can be directly used as a data source for investigating the above-mentioned research questions about exoskeletons and researchers would not have to spend time developing and/or implementing the methodology to collect relevant tweets.

### *4.2. Methodology as a Starting Point for Investigating some of the RQ's*

This section describes a methodology that may be considered as starting point for investigating some of the RQ's centered around studying the occurrence of a set of words in the tweets about exoskeletons which could include names of exoskeleton companies, specific exoskeleton products, influential people who are users of exoskeletons, politicians or other popular personalities sharing positive or negative opinions about exoskeletons. This methodology is presented to be implemented in RapidMiner [58], as RapidMiner was utilized both for the development of this dataset as well as for the statistical analysis of the same. It is also worth mentioning here that following this methodology in RapidMiner as a starting point is just one out of the many ways by which such RQ's may be investigated. This methodology would also require modifications if any of the other RQs are being investigated. Furthermore, this methodology cannot be directly applied for investigating such RQ's if RapidMiner is not being used as the application development platform. For the development of this methodology, a few extensions would have to be installed in RapidMiner Studio. These include – Text Processing 9.3 (or a higher version if available) by RapidMiner [183], Aylien Text Analysis 0.2.0 (or a higher version if available) by Aylien Inc [184], and String Matching 1.0.0 (or a higher version if available) by Aptus Data Labs [185]. These extensions are available for free in the RapidMiner Marketplace. The steps to be followed are mentioned next:

1. Hydrate all the Tweet IDs and merge the results into a single .csv file. Import this .csv file as a "*New Dataset*" into RapidMiner Studio.
2. Develop a "*Bag of Words*" model in RapidMiner Studio, which would act as a collection of keywords and/or phrases of interest related to exoskeletons. These could include names of exoskeleton companies, specific exoskeleton products, influential people who are users of exoskeletons, politicians, celebrities, or other popular personalities sharing positive or negative opinions about exoskeletons.
3. Use the "*Select Attribute*" operator to select the text of the tweets from the dataset. Identify and remove stop words from the tweet texts.

4. Use the *"Data Filters"* in RapidMiner to filter out unwanted attributes from the dataset to have only the tweets and other essential information needed for this study. Here unwanted attributes refer to non-essential details associated with each of the Tweets, such as the default profile image of the users, description mentioned on each user's profile, follower count of the users, location information of all the users, screenname of all the users, and verified status of all the user accounts, which would be obtained upon hydration using the Hydrator app.
5. Using the *"Read Document" operator*, set up a path to a file on the local system that contains a set of keywords or phrases that would be used for checking similarity and/or occurrence (Step 2). It is recommended that this file is a .txt file.
6. Implement the Levenshtein distance algorithm [186] using the *"Fuzzy matching" operator*.
7. Provide the output from the *operator* in Step 4 as the *source* and the *"Read Document" operator* (Step 5) as the grounds for comparison to generate similarity scores based on the string-comparison of each tweet
8. Enable the *advanced parameters* of the *"Fuzzy matching" operator* to define the threshold value. This can be any user-defined value, and only those tweets that have a similarity (indicating occurrence) greater than the threshold would be retained in the results
9. Integrate all the above *operators* and develop a RapidMiner *process,* and set up a Twitter connection in RapidMiner Studio.
10. Run the process to compute the results after defining specific parameters in Step 2 and Step 8.

As mentioned earlier in this Section, this step-by-step process would serve as a starting point for investigating certain RQ's only if RapidMiner is used as the application development platform. If RapidMiner is not used as the application development platform, then the following are a few resources such as libraries and/or packages in different programming languages such as Python, R, Java, C++, JavaScript, Scala, Rust, Clojure, and Ruby, that may be used to develop a similar methodology for investigating one or more of these RQ's.

- Python: Natural Language Toolkit [187], Spacy [188], TextBlob [189], Stanford Core NLP [190], PyNLPI [191], SciPY [192], Scikit-Learn [193], Keras [194], PyTorch [195], and Pandas [196].
- R: koRpus [197], OpenNLP [198], Quanteda [199], RWeka [200], Spacyr [201], Stringr [202], Text2vec [203], and Text Mining Package [204].
- Java: Apache OpenNLP [205], Apache UIMA [206], General Architecture for Text Engineering [207], LingPipe [208], Machine Learning for Language Toolkit (Mallet) [209], Natural Language Processing for JVM languages [210], and Apache Lucene [211].
- C++: MIT Information Extraction [212], MeTa [213], CRF++ [214], Colibri Core [215], InsNet [216], and Libfolia [217].
- JavaScript: Twitter Text [218], Knwl.js [219], Poplar [220], nlp.js [221], and Node Question Answering [222].
- Scala: Saul [223], ATR4S [224], Word2Vec-Sala [225], Epic [226], and TM [227].
- Rust: whatlang [228], Snips-NLU [229], and Rust-BERT [230].
- Clojure: Clojure-OpenNLP [231], Inflections-CLJ [232], and Postagga [233].
- Ruby: MonkeyLearn-Ruby [234], DialogFlow Ruby Client [235], FastText-Ruby [236], Ruby-WordNet [237], Ruby-Fann [238], TensorFlow.Rb [239], and the Ruby Language Toolkit [240].

## 5. Conclusions

The exoskeleton technology has been rapidly advancing in the last few years on account of its multitude of applications and diverse use-cases in assisted living, military, healthcare, firefighting, and industry 4.0. The exoskeleton market is projected to increase by multiple times of its current value within the next two years.

Therefore, it is crucial to study the trends of user interest, views, opinions, perspectives, attitudes, acceptance, feedback, buying behavior, and satisfaction, towards exoskeletons, for which the availability of Big Data of conversations about exoskeletons is necessary. In today's Internet of Everything era, the use of social media platforms has skyrocketed in the recent past as social media platforms provide a sense of "community" where people develop "virtual" relationships and converse on diverse topics, which includes emerging technologies such as exoskeletons. Twitter, popular amongst users of all age groups, is the second most visited social platform, and its popularity has constantly been increasing in the last few years. Researchers from different disciplines have worked on developing datasets by mining this Big Data of Twitter to record, study, interpret, and analyze conversations on Twitter related to different emerging technologies, topics, applications, matters of global concern, diseases, viruses, events, disasters, and so on; which shows the immense potential, relevance, importance, and applicability of mining of Twitter Big Data.

Even though there have been several advances in the field of exoskeleton research in the last decade and a half, no prior in this field or in the field of social media research has focused on the development of a dataset of conversations on Twitter related to exoskeletons. The work presented in this paper addresses this research challenge. It presents an open-access dataset of 138,585 Tweet IDs corresponding to 138,585 Tweets about exoskeletons posted on Twitter for a period of 5-years from May 21, 2017, to May 21, 2022. To add, based on a comprehensive review of 108 recent works in the fields of Big Data, Natural Language Processing, Information Retrieval, Data Mining, Pattern Recognition, and Artificial Intelligence that may be applied to relevant Twitter data for advancing the field of exoskeleton research, this paper presents 100 Research Questions for researchers to investigate, analyze, ideate, and explore by using this dataset. Future work would involve updating this dataset to incorporate more recent tweets as well as investigating these research questions and developing new ones to advance research and development in this field.

**Supplementary Materials:** Not applicable.

**Funding:** This research received no external funding.

**Institutional Review Board Statement:** Not applicable

**Informed Consent Statement:** Not applicable

**Data Availability Statement:** This work resulted in the creation of an open-access dataset which is available at https://dx.doi.org/10.21227/r5mv-ax79, as per the CC BY 4.0 License.

**Acknowledgments:** The author would like to thank Isabella Hall, Department of Electrical Engineering and Computer Science at the University of Cincinnati, for her assistance related to data cleaning of the first 20,000 tweets. The author would also like to thank Chia Y. Han, Department of Electrical Engineering and Computer Science at the University of Cincinnati, for his comments on improving the presentation of certain parts of the paper.

**Conflicts of Interest:** The author declares no conflict of interest.